\documentclass[%
preprint,
showpacs,preprintnumbers,
nofootinbib,
 amsmath,amssymb,
 aps,
floatfix,
]{revtex4-1}
\usepackage[colorlinks=true,
urlcolor=cambridgeblue,
linkcolor=darkraspberry,
citecolor=cambridgeblue,
linktocpage=true,
pdfproducer=medialab,
pdfa=true,
anchorcolor=blue]{hyperref}% 

\usepackage{xcolor}
\definecolor{cambridgeblue}{rgb}{0.64, 0.76, 0.68}
\definecolor{darkraspberry}{rgb}{0.53, 0.15, 0.34}

\usepackage{graphicx}% Include figure files
\usepackage{dcolumn}% Align table columns on decimal point
\usepackage{bm}% bold math
\usepackage{hyperref}% add hypertext capabilities
\usepackage{amssymb}
\usepackage{placeins}
\usepackage{float}
\usepackage[utf8]{inputenc} 
\usepackage{xcolor}

\begin{document}

\preprint{IFIC/18-25}

% \title{Displaced vertices as probes of sterile neutrinos at the high-luminosity LHC}
\title{Displaced vertices as probes of sterile neutrino mixing at the LHC}

\author{Giovanna Cottin}
\email{gcottin@phys.ntu.edu.tw}
\affiliation{Department of Physics, National Taiwan University, Taipei  10617, Taiwan}

\author{Juan Carlos Helo}
\email{jchelo@userena.cl }
  \affiliation{
Departamento de F\' isica, Facultad de Ciencias, Universidad de La Serena, 
Avenida Cisternas 1200, La Serena, Chile}
 \affiliation{ Centro-Cient\'\i fico-Tecnol\'{o}gico de Valpara\'\i so,
Casilla 110-V, Valpara\'\i so,  Chile.
}

\author{Martin Hirsch}
\email{mahirsch@ific.uv.es}
\affiliation{AHEP Group, Instituto de F\'{\i}sica Corpuscular --
    CSIC/Universitat de Val{\`e}ncia, 
    Edificio de Institutos de Paterna, Apartado 22085,
    E--46071 Val{\`e}ncia, Spain}

\date{\today}

%%%%%%%%%%%%%%%%%%%%%%%%%%%%%%%%%%%%%%%%%%%%%%%%%%%%%%%%%%%%%%%%%%%%

\begin{abstract}

We investigate the reach at the LHC to probe light sterile neutrinos
with displaced vertices. We focus on sterile neutrinos $N$ with masses
$m_{N} \sim $ (5-30) GeV, that are produced in rare decays of
the Standard Model gauge bosons and decay inside the inner trackers of
the LHC detectors. With a strategy that triggers on the prompt lepton
accompanying the $N$ displaced vertex and considers charged tracks
associated to it, we show that the 13 TeV LHC with $3000$/fb is able
to probe active-sterile neutrino mixings down to $|V_{lN}|^2\approx
10^{-9}$, with $l=e,\mu$, which is an improvement of up to four orders
of magnitude when comparing with current experimental limits from
trileptons and proposed lepton-jets searches. In the case when $\tau$
mixing is present, mixing angles as low as $|V_{\tau N}|^2 \approx 10^{-8}$ can be
accessed.

\end{abstract}

\maketitle

\section{Introduction}

Searches for new physics responsible for the lightness of neutrino
masses~\cite{Agashe:2014kda} has been in the program of the LHC
experiments for decades~\cite{Deppisch:2015qwa}. This new physics
beyond the Standard Model (SM) may be explained by the so-called
see-saw mechanism~\cite{Minkowski:1977sc} that introduces the
existence of heavy right-handed (sterile) neutrinos that mix with the
neutrinos in the Standard
Model~\cite{Mohapatra:1979ia,Schechter:1980gr}. For low enough mixing
angles and sterile neutrino masses below the electroweak scale, the
sterile neutrino $N$ can be long-lived, and may decay with a
characteristic displaced vertex (DV) signature inside particle
detectors.

Experimental efforts to search for these states at hadron colliders
normally focus on promptly decaying sterile neutrinos with masses
$m_{N}\sim\mathcal{O}(100)$ GeV. Earlier searches by
ATLAS~\cite{Aad:2015xaa} and
CMS~\cite{Khachatryan:2015gha,Khachatryan:2014dka} have focus on the
Majorana signature of same-sign dileptons and
jets~\cite{Keung:1983uu}. Only recently the CMS experiment has
provided limits for $N$ masses below $40$ GeV in the search for three
prompt charged leptons~\cite{Sirunyan:2018mtv}. For current bounds
on sterile neutrino mixing, see~\cite{Cvetic:2018elt,Das:2017nvm}. Attention to displaced
vertex signatures, which can probe masses in the GeV range, is vastly
growing. Recent phenomenological studies assessing the LHC sensitivity
with displaced vertices to light sterile neutrinos in various models
are studied
in~\cite{Deppisch:2018eth,Helo:2018qej,Jana:2018rdf,Nemevsek:2018bbt,Lara:2018rwv,Dev:2017dui}. Despite
the technical challenges in the reconstruction and modeling of the
detector response to displaced vertices, this is an important signal
of new physics as it is scarce in the Standard Model, and has to be
explored further in order to ensure the successful exploration of new
physics at the LHC, and across all mass ranges.

In this work we use a strategy motivated by the ATLAS multitrack
displaced vertex search~\cite{Aaboud:2017iio,Aad:2015rba}, recently
validated in our previous work~\cite{Cottin:2018kmq} in the context of
a left-right symmetric model. Here we focus on a simplified model in
which the Standard Model field content is extended by one heavy
sterile neutrino, briefly described in Section~\ref{sec:model}. We
consider three cases for active-sterile neutrino mixing $V_{lN}$, for 
each of the active flavours $l=e,\mu,\tau$. The
displaced vertex strategy implemented in each case is described in
Section~\ref{sec:selection}. Discovery prospects and reach at the LHC
are discussed in~\ref{sec:reach}. We close the paper and summarize in
Section~\ref{sec:summary}.

\section{Sterile neutrino simplified model}
\label{sec:model}

We consider a simplified see-saw model of neutrinos based on the
Standard Model gauge group, in which only one massive sterile neutrino
$N$ is present in the kinematic range of our interest. In this
generic framework, $N$ couples to the SM leptons via a small mixing in
the electroweak currents.  The charged and neutral current
interactions of this model are described in~\cite{Helo:2013esa}.

We consider one neutrino flavour at a time, $e,\mu$ or $\tau$,
produced in $W$ boson decays in association with the respective lepton
flavour: $W^{\pm}\rightarrow N l^{\pm}$. The $N$ decays proceed via
$N\rightarrow l^{\pm}q\bar{q}$, $N\rightarrow l'^{\mp}l^{\pm}\nu_{l}$
and $N\rightarrow \nu_{l}q\bar{q}$. The $N$ proper lifetime is given
by~\cite{Helo:2013esa}

\begin{equation}
c\tau_{N}\sim 3.7\bigg (\frac{1 \hspace{0.1cm}\mbox{GeV}}{m_{N}}\bigg)^5\bigg(\frac{0.1}{|V_{lN}|^2}\bigg)\hspace{0.2cm} [\mbox{mm}].
\label{eq:ctau}
\end{equation}

The relevant parameters are the sterile neutrino mass $m_{N}$ and
active-sterile neutrino mixing $|V_{lN}|^{2}$, which we treat as
independent. In general, small neutrino masses and mixing lead to
macroscopic lifetimes. In principle one can have a large $N$ lifetime
by making the mixing very small (instead of the sterile neutrino
mass). However, for smaller values of the mixing, the production 
rate of $N$ also becomes smaller. We are interested in masses in
the GeV range to access lifetimes of the order of picoseconds while
scanning over mixings as low as $|V_{lN}|^2=10^{-12}$.

\begin{figure}
\centering
\includegraphics[width=0.7\textwidth,angle=0]{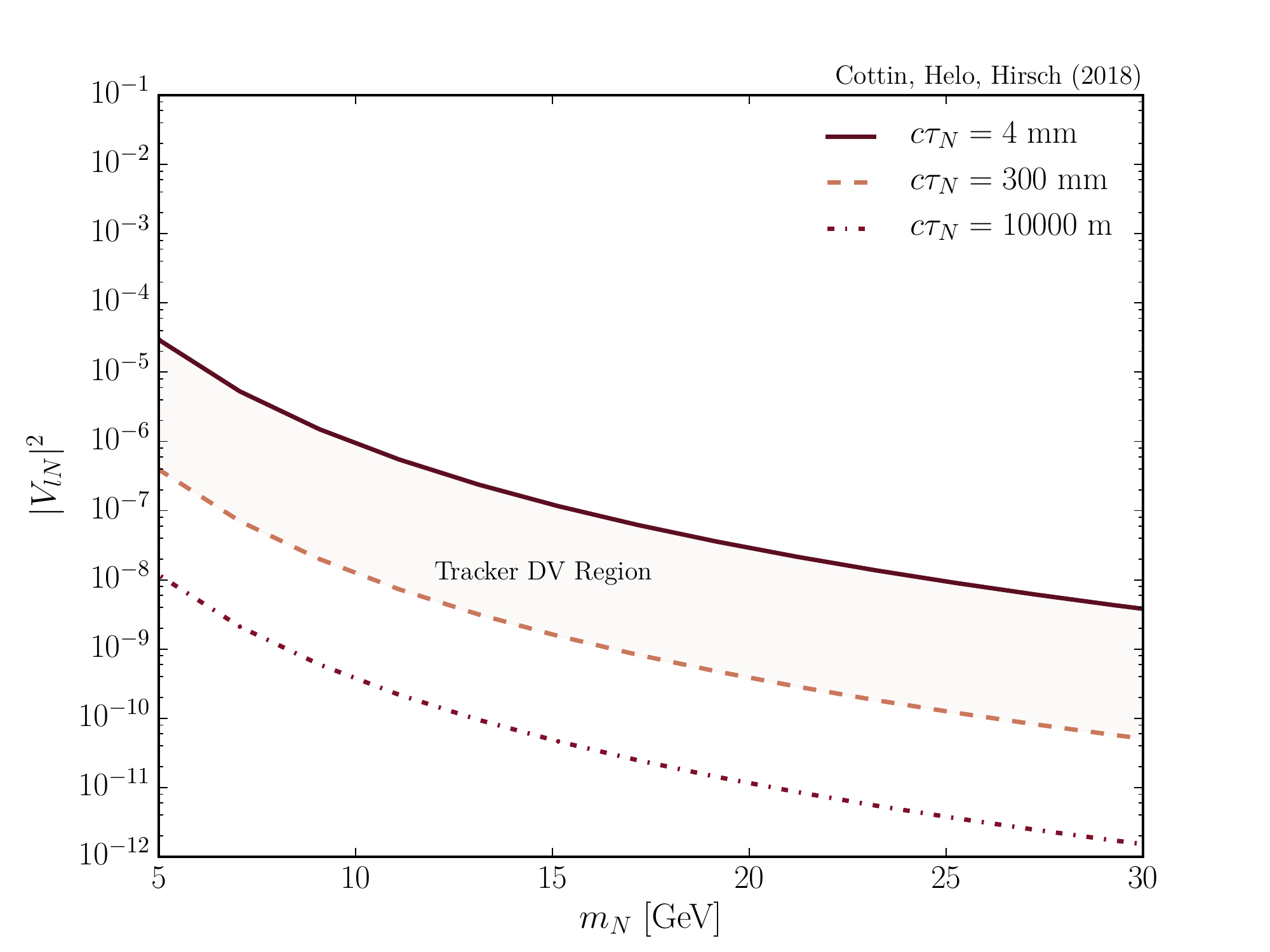}
\caption{Contours of fixed sterile neutrino proper decay distance
  $c\tau_{N}$ as a function of mass and active-sterile neutrino mixing
  angle. The shaded region represents roughly the region that can be
  accessed with current displaced vertex searches in the ATLAS inner
  tracker~\cite{Aaboud:2017iio,Aad:2015rba}. }
\label{properDecay_SST}
\end{figure}

In Figure~\ref{properDecay_SST} we show different values of the proper
lifetime in the $(m_{N},|V_{lN}|^2)$ plane. We highlight the
approximated region where vertices can efficiently be reconstructed
inside the tracker region of ATLAS, for proper decay distances between
4 and 300 mm. We are interested in sterile neutrino mass between $5$
GeV $<m_{N}<30$ GeV, which can be probed with a multitrack displaced
vertex search at ATLAS or CMS.

\section{Simulations and selection of displaced events}
\label{sec:selection}

We generate a \textsc{UFO}~\cite{Degrande:2011ua} model with
\textsc{SARAH}~\cite{Staub:2013tta} and use
\textsc{SPheno}~\cite{Porod:2011nf,Porod:2003um} for the spectrum
calculation of the sterile neutrino simplified model. We simulate
events at $\sqrt{s}=13$ TeV for the process $pp\rightarrow
W^{\pm}\rightarrow N l^{\pm}$. Generation is performed with
\textsc{MadGraph5\_aMC@NLOv2.4.3}~\cite{Alwall:2014hca} at leading
order. We normalize the corresponding value to match the experimental
cross section in Ref~\cite{Aad:2016naf}. The generated events are then
interfaced to \textsc{Pythia8 v2.3}~\cite{Sjostrand:2014zea} for
showering, hadronization and computation of the $N$ decays. Plots are
generated with {\texttt{matplotlib}}~\cite{Hunter:2007}.

The promising decay channels for the sterile neutrino are
semileptonically $N\rightarrow l^{\pm}q\bar{q}$ and leptonically
$N\rightarrow l'^{\mp}l^{\pm}\nu_{l}$. Decays via a neutral current
such that $N\rightarrow \nu_{l}q\bar{q}$ are also
possible~\cite{Helo:2013esa}. All these modes will lead to displaced
vertices with charged tracks associated to them. For the case when
only $\tau$ mixing is present, both semileptonic and leptonic decays
lead to the presence of a $\tau$ lepton coming from the displaced
vertex.

We propose a search inspired by the ATLAS multitrack displaced vertex
analysis~\cite{Aaboud:2017iio,Aad:2015rba}, which is sensitive to
lifetimes of the order of picoseconds to about a nanosecond, so
particle decays can be reconstructed with a displaced vertex
signature inside the inner tracker. This strategy was developed in our
recent work in Ref.~\cite{Cottin:2018kmq}, where we trigger on the
prompt lepton coming from the $W$ boson decay, impose cuts on the
neutrino displaced vertex and its decay products, and apply
vertex-level efficiencies (made public by ATLAS
in~\cite{Aaboud:2017iio}) to DVs that pass the required particle-level
acceptance cuts. Since the parametrized selection efficiencies
provided by ATLAS assumes all decay products are prompt from the DV,
they are not directly applicable to the case when there is a $\tau$
lepton coming from the displaced vertex~\footnote{The further displacement of
  $\tau$'s inside the vertex will affect the vertex reconstruction
  efficiency from the subsequent displacement of taus (and from any
  heavy flavour quark in general). This was addressed for example in
  Ref.~\cite{Allanach:2016pam}, when there are two $b'$s coming from
  the displaced vertex. By allowing a bigger merging distance of $5$
  mm (instead of $1$ mm) when forming a vertex, some efficiency is
  recovered. Since we do not implement a vertex reconstruction
  algorithm in this work, the loss in efficiency can not be
  estimated.}, so we consider decays to $N\rightarrow \nu_{l}q\bar{q}$
only when $\tau$ mixing is present.

Prompt leptons are reconstructed considering the following:

{\it{For electrons:}} We require an isolated electron within
$|\eta|<2.5$. We smear their momenta with a resolution of $2\%$ at 10
GeV, falling linearly to $1\%$ at 100 GeV, and then $1\%$ flat.

{\it{For muons:}} We require an isolated muon within $|\eta|<2.5$.  We
smear their momenta with a resolution between $|\eta|$ of $2$ and $0$,
linearly falling from $4\%$ to $1.5\%$.

{\it{For taus:}} We implement a basic reconstruction following
Ref.~\cite{Aad:2014rga}. We start by reconstructing jets with
{\textsc{FastJet 3.1.3}}~\cite{Cacciari:2011ma} using the anti$-k_{t}$
algorithm with distance parameter $R = 0.4$. Only jets with $p_{T}>10$
GeV and $|\eta|<2.5$ are taken as seeds for $\tau$
reconstruction. Charged constituents inside the jet must have
$p_{T}>1$ GeV. If a truth $\tau$ candidate falls within a cone $\Delta
R<0.2$ centered on the jet axis, it is selected.

The following selections are then imposed:

\begin{enumerate}

\item{One prompt lepton (as reconstructed above) with $p_{T}>25$ GeV.}

\item{Decay position of the DV contained within transverse distance
  $r_{DV}<300$ mm, and $|z_{DV}|<300$ mm. The distance between the
  interaction point and the decay position must be bigger than $>4$
  mm.}

\item{Decay products must be charged (i.e tracks) with $p_{T}>1$ GeV
  and transverse impact parameter $|d_{0}|>2$ mm. $d_{0}$ is defined
  as $d_{0}=r_{DV}\times \sin{\Delta\phi}$, with $\Delta\phi$ being
  the azimuthal angle between the decay product and the trajectory of
  the long-lived $N$.}

\item{ The number of selected tracks $N_{trk}$ must be at least 3. The invariant
  mass of the DV $m_{DV}$ must be $\geq5$ GeV, and assumes all tracks have the
  mass of the pion.}

\item{Parametrized selection efficiencies are applied depending on the
  displaced vertex distance (within 4 and 300 mm, between the pixel
  and the SCT), number of tracks and mass.}
\end{enumerate}

As pointed out in~\cite{Cottin:2018kmq}, with these selections we are
still in a zero background region, where background comes mostly from
instrumental sources.

\section{Sensitivity reach}
\label{sec:reach}

We analyze the region where a displaced search with the above
selections can have sensitivity. The relevant parameters in the
sterile neutrino model are the neutrino mass $m_{N}$ and
active-sterile neutrino mixing $|V_{lN}|^{2}$.

We first chose a representative benchmark point with $m_{N}=15$ GeV, $|V_{lN}|^2=10^{-8}$
and proper neutrino decay distance $c\tau_{N} \approx 50$ mm to illustrate the effect of all analysis 
cuts from the previous section. We combine the second and third cuts in the ``DV fiducial'' entry 
in Table~\ref{tab:sens}. Events are generated at $13$ TeV which corresponds to a production cross section 
of $0.09$ fb. Cut-flows in the case of electron, muon and tau mixing are shown.

As already noted in the context of a left-right symmetric model
in~\cite{Cottin:2018kmq}, for sterile neutrinos with masses below the
electroweak scale this search strategy looses sensitivity, as lower
masses will lead to softer decay products with limited amount of
tracks available to make up a vertex. However, given the low
background nature of this signature, the discovery of a
displaced vertex signal in the sterile neutrino model is
possible at the high-luminosity LHC.

\begin{table}[h]
\begin{center}
\begin{tabular*}{0.65\textwidth}{@{\extracolsep{\fill}}  l  c  c  c }
\hline
                     &$e$ mixing         &   $\mu$ mixing       &    $\tau$ mixing      \\
                     &$\epsilon=4.88 \%$ &   $\epsilon=4.21 \%$  &    $\epsilon=0.7 \%$      \\

\hline \hline 
All events          & 10000  & 10000  & 10000  \\
Prompt $l$          &  4004  &  3321  & 764     \\
DV fiducial         &  2266  &  2058  & 433     \\
$N_{trk}$           &  814   &    712 &  177    \\
$m_{DV}$            &  645   &    579 &  72     \\
DV efficiency       &  488   &    421 &  70      \\
\hline 
\end{tabular*} 
\caption{Numbers of simulated events at $\sqrt{s}=13$ TeV for a benchmark 
with $m_{N}=15$ GeV, $|V_{lN}|^2=10^{-8}$, and $c\tau_{N}\approx 50$ mm, in the three cases 
considered where $l= e , \mu$ or $\tau$. Overall efficiencies $\epsilon$ after all cuts are also shown.}
\label{tab:sens}
\end{center}
\end{table}

\begin{figure}
\centering
\includegraphics[width=0.6\textwidth,angle=0]{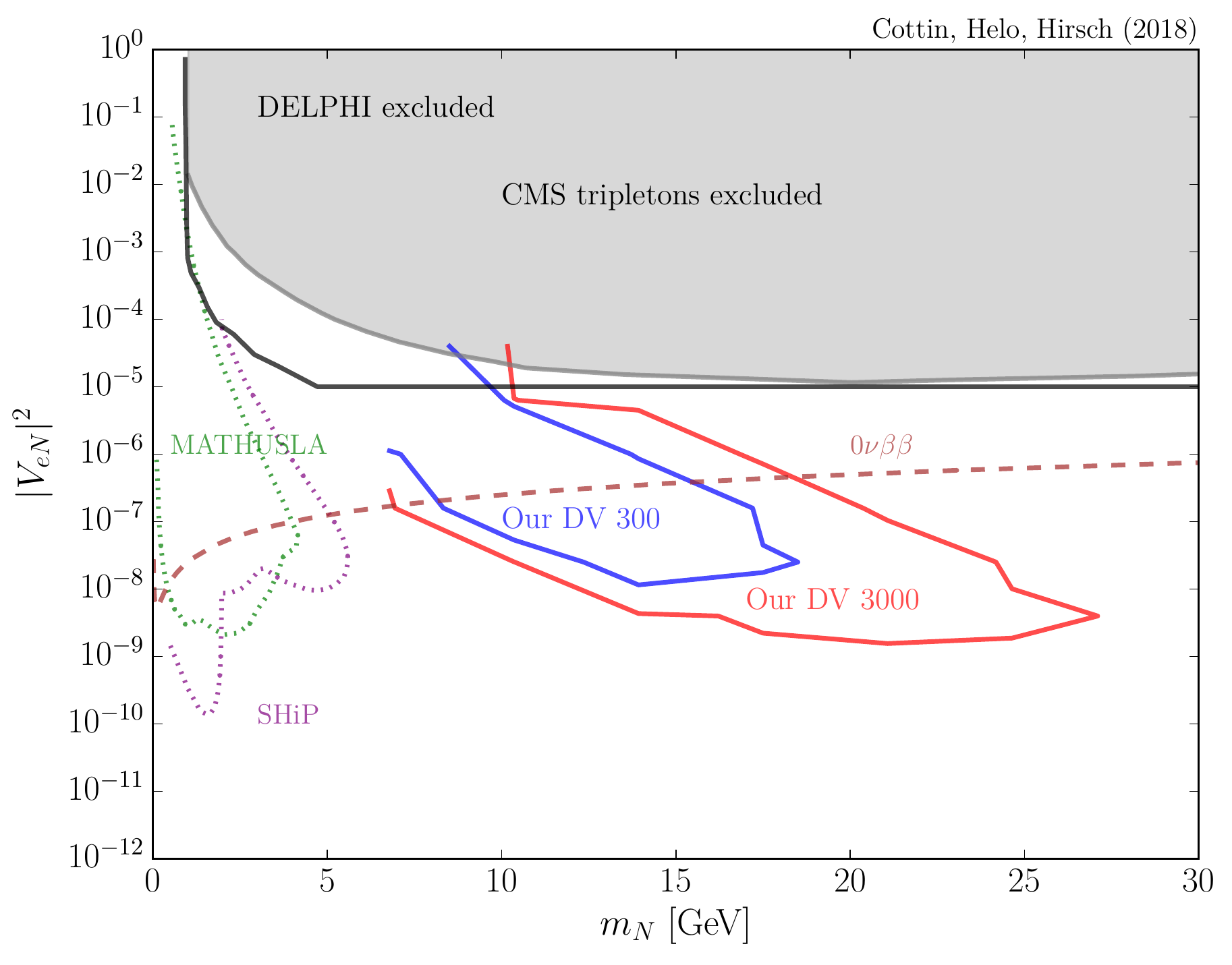}
\caption{$95\%$ CL reach in the $(|V_{eN}|^2,m_{N})$ plane at
  $\sqrt{s}=13$ TeV of our proposed multitrack displaced strategy for
  $\mathcal{L}=300$ fb$^{-1}$ (blue) and $\mathcal{L}=3000$ fb$^{-1}$
  (red). Projected sensitivities for the MATHUSLA (dashed green) and SHiP (dashed purple) experiments are also shown for comparison
  (taken from Ref.~\cite{Helo:2018qej} and~\cite{Bondarenko:2018ptm}, respectively). The DELPHI limit, taken from~\cite{Deppisch:2015qwa},
  is also shown. A limit derived from $0\nu\beta\beta$ experiments is
  also shown (see text for more details). The $95\%$ CL exclusion of
  the CMS 13 TeV trileptons search~\cite{Sirunyan:2018mtv} is given by
  the filled grey region.}
\label{reach_SST_e}
\end{figure}

\begin{figure}
\centering
\includegraphics[width=0.6\textwidth,angle=0]{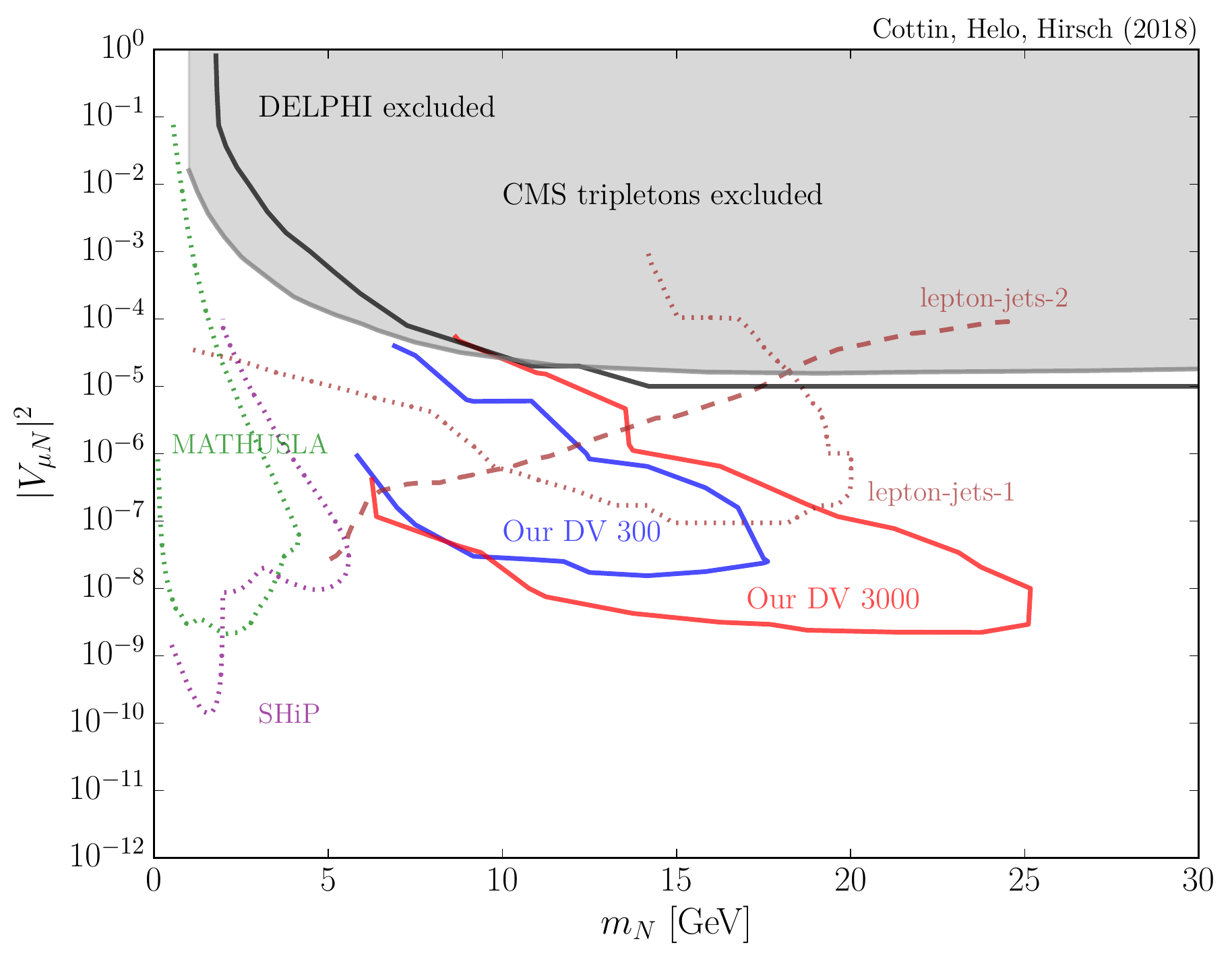}
\caption{$95\%$ CL reach in the $(|V_{\mu N}|^2,m_{N})$ plane at
  $\sqrt{s}=13$ TeV of our proposed multitrack displaced strategy for
  $\mathcal{L}=300$ fb$^{-1}$ (blue) and $\mathcal{L}=3000$ fb$^{-1}$
  (red). Projected sensitivities for the MATHUSLA (dashed green) and SHiP (dashed purple) experiments are also shown for comparison
  (taken from Ref.~\cite{Helo:2018qej} and~\cite{Bondarenko:2018ptm}, respectively).  The DELPHI limit, taken
  from~\cite{Deppisch:2015qwa}, is also shown. The 13 TeV limit from
  proposed lepton-jets searches are shown in dashed brown, taken from
  Ref.~\cite{Izaguirre:2015pga} (``lepton-jets1") and
  Ref.~\cite{Dube:2017jgo} (``lepton-jets2"). See the text for more
  details. The $95\%$ CL exclusion of the CMS 13 TeV trileptons
  search~\cite{Sirunyan:2018mtv} is given by the filled grey region.}
\label{reach_SST_mu}
\end{figure}

\begin{figure}
\centering
\includegraphics[width=0.6\textwidth,angle=0]{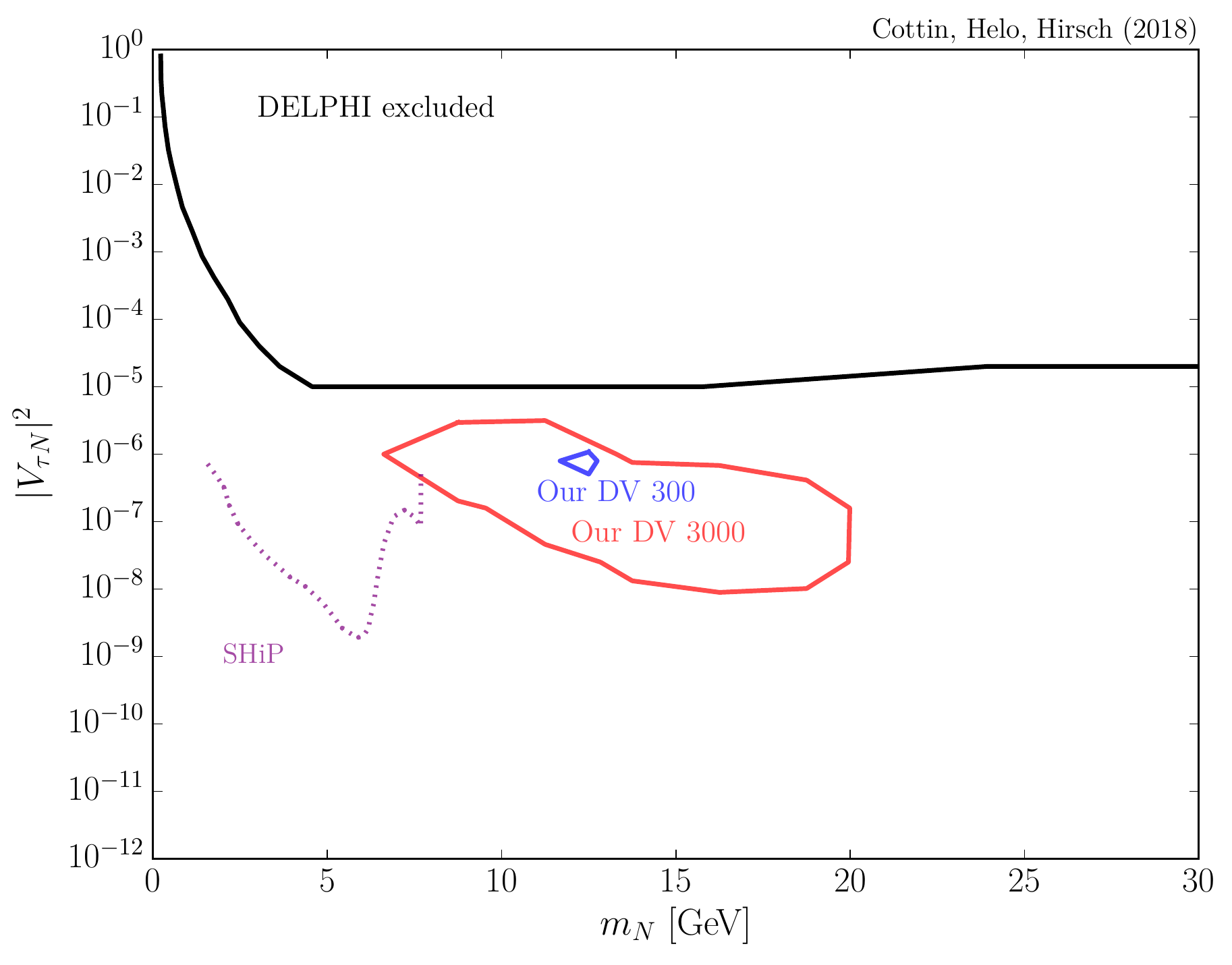}
\caption{$95\%$ CL reach in the $(|V_{\tau N}|^2,m_{N})$ plane at
  $\sqrt{s}=13$ TeV of our proposed multitrack displaced strategy for
  $\mathcal{L}=300$ fb$^{-1}$ (blue) and $\mathcal{L}=3000$ fb$^{-1}$
  (red). Projected sensitivity the MATHUSLA (dashed green) and for the SHiP (dashed purple) experiment
  is also shown for comparison (taken from Ref.~\cite{Curtin:2018mvb}). The DELPHI limit, taken
  from~\cite{Deppisch:2015qwa}, is also shown for comparison.}
\label{reach_SST_tau}
\end{figure}

Figures~\ref{reach_SST_e},~\ref{reach_SST_mu} and~\ref{reach_SST_tau}
show the estimated reach in the $(m_{N},|V_{lN}|^2)$ plane, 
in the case of electron, muon and tau mixing, respectively. We also show for 
comparison the projected sensitivities from
the proposed SHiP~\cite{Alekhin:2015byh} and MATHUSLA~\cite{Curtin:2018mvb,Chou:2016lxi} 
experiments. For projected sensitivities for
other proposals for heavy neutral 
leptons within $m_{N}<5$ GeV, see FASER~\cite{Feng:2017uoz,Kling:2018wct,Helo:2018qej} 
and CODEX-b~\cite{Gligorov:2017nwh,Helo:2018qej}.

In Figure~\ref{reach_SST_e} we see that mixings as low
as $|V_{e N}|^2\approx 1.5 \times 10^{-9}$ for $3000/$fb and 13 TeV
can be probed for $m_{N}=20$ GeV. Current neutrinoless double beta
decay ($0\nu\beta\beta$) experiments and searches for trileptons at
the LHC are also sensitive to $\mathcal{O}(10)$ GeV sterile neutrino
masses. We show an update of the limits calculated
in~\cite{Helo:2013esa} using the latest limit on $0\nu\beta\beta$ from
the GERDA experiment~\cite{Agostini:2018tnm}.  LHC limits for sterile
neutrino masses below $40$ GeV are presented for the first time in the
CMS 13 TeV search for three prompt charged leptons in the final
state~\cite{Sirunyan:2018mtv}. Other significant constrain in our mass
region of interest comes from LEP data, where the DELPHI collaboration
provides limits on sterile states produced in decays of the $Z$
boson~\cite{Abreu:1996pa}.

In Figure~\ref{reach_SST_mu}, mixings as low as $|V_{\mu N}|^2\approx
2.2 \times 10^{-9}$ for $3000/$fb and 13 TeV can be probed for
$m_{N}=20$ GeV. We also show limits from proposals with lepton-jets
searches, which is a complementary strategy. The curve labeled
``lepton-jets-1" shows the exclusion form
Ref.~\cite{Izaguirre:2015pga} at 13 TeV and $300/$fb, where zero
background is assumed. The curve labeled ``lepton-jets-2" shows the 13
TeV limit in~\cite{Dube:2017jgo} and $300/$fb, and considers
additional background sources to the ones
in~\cite{Izaguirre:2015pga}. An improvement of roughly two orders of
magnitude in sensitivity is achieved with our strategy with $300/$fb
for masses $5<m_{N}<25$ GeV. The $95\%$ CL exclusion of the CMS 13 TeV
prompt trileptons search~\cite{Sirunyan:2018mtv} is also shown,
proving to be competitive with DELPHI~\cite{Abreu:1996pa}.

Finally, we show in Figure~\ref{reach_SST_tau} the reach when there is
$\tau$ mixing. Experimental limits for $\tau$ mixing have not been
addressed yet at the LHC. We show the DELPHI limit~\cite{Abreu:1996pa}
from $Z$ decays for comparison. For taus, a large luminosity sample will 
be needed for obtaining meaningful constraints. As the figure shows there 
is only a very narrow region testable with $300$/fb. Mixings as low 
as $|V_{\tau N}|^2 \approx 8\times 10^{-9}$ for $m_{N}=16$ GeV at $3000$/fb 
can be probed with this strategy.

\section{Summary and Conclusions}
\label{sec:summary}

We study the potential of the LHC to probe light sterile neutrinos,
and active-sterile neutrino mixing angles, with a displaced vertex
strategy motivated by current multitrack DV searches at the 13 TeV
LHC. We focus on a simplified model where the Standard Model is
extended with one sterile neutrino $N$. Mixing with the three flavours
$e,\mu$ and $\tau$, are treated separately. In all cases, to our
knowledge, we see that this strategy probes to be the most sensitive
to date in the mass region of interest ($5$ GeV $<m_{N}<30$ GeV).

For $e$ mixing, we show that this DV search is more sensitive than
current neutrinoless double beta decay experiments. In the case of
$\mu$ mixing, parts of the parameter space not accessible with other
LHC searches (such as lepton-jets or trileptons) is possible. In both
cases, with $3000$/fb, an improvement of up to four orders of
magnitude ($|V_{l N}|^2\approx \times 10^{-9}$ for sterile neutrino
masses between $5$ and $20$ GeV) in sensitivity is gained when
comparing with the current experimental limits from trileptons
searches at CMS~\cite{Sirunyan:2018mtv}.

Accessing $\tau$ mixing is less straightforward due to to the
difficulty in reconstructing $\tau$ leptons, and also since the
presence of a subsequently displaced $\tau$ coming from the displaced
vertex affects the vertex reconstruction efficiency in a way that is
not trivial to quantify. We access $\tau$ mixing by considering
sterile neutrino decays via a neutral current only, which still leads
to a displaced vertex formed from hadronized tracks, to which 
publicly available DV efficiencies can be applied, thus
avoiding the problematic $\tau$ in the DV. This may be an advantage of
a multitrack based strategy, as opposed to tagging leptons coming from
the DV, in constraining $\tau$ mixing.

We briefly comment on prospects for testing sterile neutrinos at
LHCb. The authors in~\cite{Antusch:2017hhu} discuss limits for a
different model than the one presented in this work. They show that
sterile neutrino masses around $9$ GeV and mixings down to $\approx
10^{-6}$ can be constrained in semileptonic sterile neutrino decays
($\mu q \bar{q}$) at the $95\%$ CL with current LHCb
data~\cite{Aaij:2016xmb}. Mixings up to $\approx 10^{-8}$ (for masses
between $15-25$ GeV) can be further probed at higher luminosity,
suggesting limits for $\mu$ mixing from LHCb could be competitive in
this mass region to the ones we derived here in the context of ATLAS.

Finally, the sensitivity of this displaced strategy at the LHC is
complementary to that of future fixed-target experiments, such as
SHiP, or the MATHUSLA surface detector, which can probe sterile
neutrino masses below 5 GeV. This makes a tracker based DV search for
light sterile neutrinos unique, as it has no competition from other
experiments within $5$ GeV $<m_{N}<30$ GeV.

\acknowledgments{G.C. acknowledges support by the Ministry of Science and
  Technology of Taiwan under grant No. MOST-106-2811-M-002-035. J.C.H.
  is supported by Chile grants Fondecyt No. 1161463, Conicyt PIA/ACT
  1406 and Basal FB0821. M. H. was funded by Spanish MICINN grant
  FPA2017-85216-P and SEV-2014-0398 (from the Ministerio de Economía,
  Industria y Competitividad), as well as PROMETEOII/2014/084 (from
  the Generalitat Valenciana).}

\bibliographystyle{apsrev4-1}
\bibliography{main}% Produces the bibliography via BibTeX.

\end{document}